\documentclass[prb,twocolumn,showpacs,preprintnumbers,amsmath,amssymb]{revtex4}

\usepackage{graphicx}
\usepackage{bbm}

\begin{document}
\preprint{}
\title{First principles investigation of ferroelectricity in epitaxially strained Pb$_2$TiO$_4$}
\author{Craig J. Fennie and Karin M. Rabe}
\affiliation{Department of Physics and Astronomy, Rutgers University,
        Piscataway, NJ 08854-8019}
\date{\today}

\begin{abstract}
The structure and polarization of the as-yet hypothetical Ruddlesden-Popper compound 
Pb$_2$TiO$_4$ are investigated within density-functional theory. Zone center phonons 
of the high-symmetry K$_2$NiF$_4$-type reference structure, space group $I4/mmm$, 
were calculated. At the theoretical ground-state lattice constants, there is one 
unstable infrared-active phonon. This phonon freezes in to give the $I2mm$ ferroelectric 
state. As a function of epitaxial strain, two additional ferroelectric phases are found, 
with space groups $I4mm$ and $F2mm$ at compressive and tensile strains, respectively.

\end{abstract}


\maketitle

%%%%%%%%%%%%%%%%%%%%%%%%%%%%%%%%%%%%%%%%%%%%%%%%%%%%%%%%%%%%%%%%%%%%%%%%%%%%%%%%%%
%INTRODUCTION
%%%%%%%%%%%%%%%%%%%%%%%%%%%%%%%%%%%%%%%%%%%%%%%%%%%%%%%%%%%%%%%%%%%%%%%%%%%%%%%%%%
%\section{Introduction}
%\label{sec:intro}

The occurrence of ferroelectricity in the $AB$O$_3$ perovskite structure
has been known since the 1950's.~\cite{lines.glass} Recently, first-principles 
density functional methods have proved invaluable in elucidating the observed 
behavior of known perovskite oxide ferroelectrics, anti-ferroelectrics, and 
quantum paraelectrics.~\cite{ghosez.prb.99,cohen.jpcs.00} Examples include the 
alkaline-earth titanates,~\cite{cohen.nature.92,lasota.ferro.97,cockayne.prb.00} 
the alkali metal tantalates and niobates,~\cite{singh.prb.96,inbar.prb.96} and the 
lead-based perovskites.~\cite{cohen.nature.92,singh.prb.95} There is also a growing 
interest in applying these methods to the design of new ferroelectrics based on 
the perovskite structure.~\cite{spaldin.ssc.03,halilov.apl.02}

\begin{figure}[b]
\includegraphics[scale=0.1]{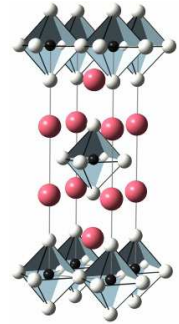}\\
\caption{\label{fig:RP_I4mmm} (color online) Crystal structure of Ruddlesden-Popper
compound Pb$_2$TiO$_4$, Space group $I$4$/mmm$.}
\end{figure}

Another path to new materials leads beyond the perovskite structure.
The Ruddlesden-Popper (RP) family of compounds are closely related to the 
perovskites.~\cite{ruddlesden.both} They can be viewed as a stacking 
$A$O-terminated $AB$O$_3$ perovskite [001] slabs of thickness equal to $n$ 
primitive lattice constants. Adjacent slabs are shifted relative to one 
another along [110] by $a$/2, giving the homologous series $A_{n+1}B_n$O$_{3n+1}$. 
The $n$=1 structure is shown in Fig.~\ref{fig:RP_I4mmm}.

Given the structural similarity to the perovskites, it seems surprising that 
there have been no confirmed cases of ferroelectricity in the RP family of 
compounds.~\cite{my.Ca3Ti2O7} Bulk phases of RP titanates have been reported 
only for some members of the Sr$_{n+1}$Ti$_n$O$_{3n+1}$ and Ca$_{n+1}$Ti$_n$O$_{3n+1}$ 
series.~\cite{my.KLnTiO4} It should not be surprising that neither the strontium 
nor the calcium RP series of compounds appear to display ferroelectricity given 
that the end members ($n$=$\infty$) are SrTiO$_3$ and CaTiO$_3$, respectively, 
neither of which themselves are ferroelectric. Still, the fact that many RP titanates 
are not thermodynamically stable does not preclude the possibility that a metastable 
RP ferroelectric phase could be produced by an appropriate 
synthetic process. In order to identify such a system, it is convenient to use 
first-principles density functional (DFT) methods, for example, to investigate an as-yet 
hypothetical first member of a RP series (where the effects of the structural 
modification would be most severe) whose end member is a perovskite ferroelectric.

In this paper we show that Pb$_{2}$TiO$_{4}$, an $n$=1 RP compound based on 
ferroelectric end-member PbTiO$_3$, is a promising candidate for a high-polarization 
ferroelectric. Regarding the presence of Pb$_{2}$TiO$_{4}$ (or higher $n$ compounds) 
in the bulk phase diagram, reports are conflicting and no structural information is 
available.~\cite{belyaev.rjic.70,eisa.tjbcs.80,soh.calphad.94} However, as discussed 
above, it may be possible to produce a metastable form using modern epitaxial growth 
techniques. Using first-principles DFT calculations, we compute the ground-state 
structure and polarization, finding that Pb$_{2}$TiO$_{4}$ is a ferroelectric with a 
polarization comparable to PbTiO$_3$. Furthermore, the direction of the polarization 
can be changed by an applied epitaxial strain.

%%%%%%%%%%%%%%%%%%%%%%%%%
% Figure 1
%%%%%%%%%%%%%%%%%%%%%%%%%%

%%%%%%%%%%%%%%%%%%%%%%%%%%%%%%%%%%%%%%%%%%%%%%%%%%%%%%%%%%%%%%%%%%%%%%%%%%%%%%%%%%%
%METHOD
%%%%%%%%%%%%%%%%%%%%%%%%%%%%%%%%%%%%%%%%%%%%%%%%%%%%%%%%%%%%%%%%%%%%%%%%%%%%%%%%%%%%
%\section{Method}
%\label{sec:method}
First-principles DFT calculations were performed within
the local density approximation as implemented in the Vienna {\it ab initio} 
Simulations Package (VASP).~\cite{VASP} A plane wave basis set and 
projector-augmented wave potentials were employed.~\cite{PAW} 
The electronic wavefunctions were expanded in plane waves up to a kinetic energy 
cutoff of 500 eV.  Integrals over the Brillouin zone were approximated by sums on 
a $6 \times 6 \times 6$ $\Gamma$-centered mesh of $k$-points. Polarization was 
calculated using the modern theory of polarization as implemented in 
VASP.~\cite{king-smith.prb.93}

We approach the problem of searching for possible ferroelectric states by 
first calculating the properties of the RP high-symmetry reference structure,
space group: $I4/mmm$, the structure one expects for the 
paraelectric phase. We performed full optimization of the lattice parameters 
($a$=3.857\AA, $c$=12.70\AA) and internal coordinates ($z_{Pb}$=0.3507, 
$z_{O_{\mathrm{II}}}$=0.1562). The residual Hellmann-Feynman forces were 
less than 2 meV/\AA. Next we calculated the zone-center phonons of this reference 
system by computing the dynamical matrix at $q=0$ using the direct method where 
each ion was moved by approximately 0.01$\AA$. We then froze in the real-space 
eigendisplacements of selected unstable modes and performed full relaxations in 
the space group determined by the symmetry breaking mode. Finally, we compute the 
polarization.

%For analysis of the effects of epitaxial strain, we constrained the two basal 
%primitive vectors of the bct lattice to an angle of 90 degrees and 
%equal length corresponding to that of an implicit square-lattice substrate. 
%Calculations of the ground state $I4/mmm$ structures, ferroelectric instabilities 
%and energy, structure and polarization of low-symmetry phases were carried out as 
%in the previous paragraph.

%%%%%%%%%%%%%%%%%%%%%%%%%%%%%%%%%%%%
% Table 1
\begin{table}[t]
\caption{Crystal structure of ferroelectric Pb$_2$TiO$_4$,
Space Group: $I2mm$, $a$=3.985\AA, $b$=3.826\AA, $c$=12.70\AA.}
\begin{ruledtabular}
\begin{tabular}{lcl}
Atom &Wyckoff&Coordinates\\ \hline
\begin{tabular}{l}Pb\\Ti\\O$_{\mathrm{Ix}}$\\
O$_{\mathrm{Iy}}$\\O$_{\mathrm{II}}$\\\end{tabular}
&\begin{tabular}{ll} (4c)&$\,\,\,m$\\(2a)&$2$$mm$\\
(2a)&$2$$mm$\\(2b)&$2$$mm$\\(4c)&$\,\,\,m$\\\end{tabular}
&\begin{tabular}{lllll} \,\,0.0616&&0&&0.3508\\
\,\,0.0256&&0&&0\\-0.0190+$1\over2$&&0&&0\\
-0.0307&&$1\over2$&&0\\-0.0496&&0&&0.1538\\ \end{tabular}
\end{tabular}
\end{ruledtabular}
\label{table:struct.I2mmFull}
\end{table}
%%%%%%%%%%%%%%%%%%%%%%%%%%%%%%%%%%%%%

%%%%%%%%%%%%%%%%%%%%%%%%%%%%%%%%%%%%%%%%%%%%%%%%%%%%%%%%%%%%%%%%%%%%%%%%%%%%%%%%%%%%
%RESULTS
%%%%%%%%%%%%%%%%%%%%%%%%%%%%%%%%%%%%%%%%%%%%%%%%%%%%%%%%%%%%%%%%%%%%%%%%%%%%%%%%%%%5
%\section{Results and Discussion}
%\label{sec:results}

%%%%%%%%%%%%%%%%%%%%%%%%%%%%%%%%%%%%%%%%%%%%%%%%%%%%%%%%%%%%%%%%%%%%%%%%%%%%%%%%%%%%
%\subsection{Ferroelectric Pb$_2$TiO$_4$ ground state}

%\label{sec:results.gs}

For Pb$_2$TiO$_4$ in the $I4/mmm$ high-symmetry reference structure, 
there are three infrared-active (i.r.) modes that transform according 
to the irreducible representation $A_{2u}$ and four i.r. modes that 
transform according to $E_{u}$. The one-dimensional (1-d) $A_{2u}$ modes 
involve atomic distortions along [001] while in the 2-d $E_{u}$ modes, 
atoms move in the plane perpendicular to [001]. Our calculations reveal 
that at the ground-state structural parameters, $I4/mmm$ Pb$_2$TiO$_4$ 
has one phonon with an imaginary frequency ($\omega$=96$i$ cm$^{-1}$),
indicative of an instability. This unstable phonon is infrared-active 
of E$_u$ symmetry type. Therefore, {\it Pb$_2$TiO$_4$ in the RP structure 
does indeed display a ferroelectric instability}. The real-space 
eigendisplacements of this unstable ferroelectric mode consists of Pb and 
Ti atoms moving against a non-rigid oxygen octahedra (with larger displacements 
of the apical oxygens in the PbO layer). As for PbTiO$_3$,~\cite{ghosez.prb.99} 
the character of the ferroelectric instability in Pb$_2$TiO$_4$ involves 
significant Pb displacements moving against oxygen in the Pb-O planes. This 
involvement of the $A$-site cation in both PbTiO$_3$ and Pb$_2$TiO$_4$ differs 
from non-Pb based compounds (e.g. BaTiO$_3$) and has been attributed to the 
Pb$^{2+}$ 6s$^2$ lone-pair. It is in fact this lone-pair physics that stabilizes 
PbTiO$_3$ in the tetragonal phase~\cite{cohen.nature.92} and may have a role 
in facilitating the ferroelectric distortion in Pb$_2$TiO$_4$.~\cite{seshadri.chemmat.01}

The key role of Pb is further emphasized by comparison with Ba$_2$TiO$_4$. 
A compound at this composition, barium orthotitanate, has been identified in 
the bulk phase diagram. It crystallizes not in the RP structure but rather in 
the monoclinic distorted-K$_2$SO$_4$ structure (space group: P2$_1$/n).~\cite{bland.acta.61} 
A calculation for the structure and zone-center phonons of  $I4/mmm$ Ba$_2$TiO$_4$ 
in the hypothetical RP structure, exactly analogous to that for Pb$_2$TiO$_4$, 
shows no evidence for any ferroelectric instability, even for varying epitaxial strain.

Returning to Pb$_2$TiO$_4$, we search for the ferroelectric ground state by 
freezing-in the real-space eigendisplacement pattern of the unstable E$_u$ mode,
and performing full relaxations of all ions and lattice constants consistent
with the resultant space group. Since this E$_u$ mode is doubly degenerate,
any linear combination of the degenerate modes polarized along [100] and 
[010] is an equivalent eigendisplacement. We considered two linear combinations; 
one polarized along [100], a second polarized along [110]. Freezing-in the E$_u$ 
mode polarized along [100] results in a body-centered orthorhombic space group, 
$I2mm$. For the distortion along [110], the resulting space group is face-centered 
orthorhombic  $F2mm$. Since this $F2mm$ structure is slightly higher in energy 
(20 meV/formula unit) than the $I2mm$ structure, we now consider only the latter 
(we will revist $F2mm$ below when discussing the effect of epitaxial strain). Our 
calculated structural parameters of Pb$_2$TiO$_4$ in this orthorhombic space group 
are displayed in Table~\ref{table:struct.I2mmFull}. We imposed the convention, 
$\sum_i \Delta x_i$=0, where $\Delta x_i$ is the deviation along [100] from the 
centrosymmetric position of ion $i$. It can be seen that the relaxed structure can 
in fact be related to the high-symmetry RP structure by small displacements of Pb 
and Ti ions moving against the non-rigid oxygen octahedra, consistent with the 
freezing-in of the E$_u$ phonon instability of the high-symmetry structure as 
proposed. Finally, we calculate the spontaneous polarization P$_s$. We find that 
P$_s= 68 \mu$C/cm$^2$, along [100] as required by symmetry. 
Therefore {\it Pb$_2$TiO$_4$ in the RP structure is a ferroelectric with 
a spontaneous polarization comparable to that of PbTiO$_3$}.~\cite{my.pbtio3}

%%%%%%%%%%%%%%%%%%%%%%%%%%%%%%%%%%%%%%%%%%%%%%%%%%%%%%%%%%%%%%%%%%%%%%%%%%%%%%%%%%%%
%\subsection{Epitaxially stabilized ferroelectric phases}
%\label{sec:epitax}
Epitaxy plays a dual role in our thinking about the Pb$_2$TiO$_4$ system.
As will be discussed shortly, one possible route to 
synthesize thin films of Pb$_2$TiO$_4$ in the RP structure is through 
the use of epitaxial stabilization.~\cite{schlom.matsci.01,gorbenko.chemmat.02} 
In addition, it is becoming increasingly possible to grow oxide thin films
coherently on substrates with a relatively wide range of lattice constants
(1-2$\%$ lattice mismatch is currently the norm). This provides an 
additional parameter to ``tune'' the properties of the material to desired values 
by applying an in-plane (or epitaxial) strain to the thin film compared to bulk.

\begin{figure}[t]
\includegraphics[scale=0.25]{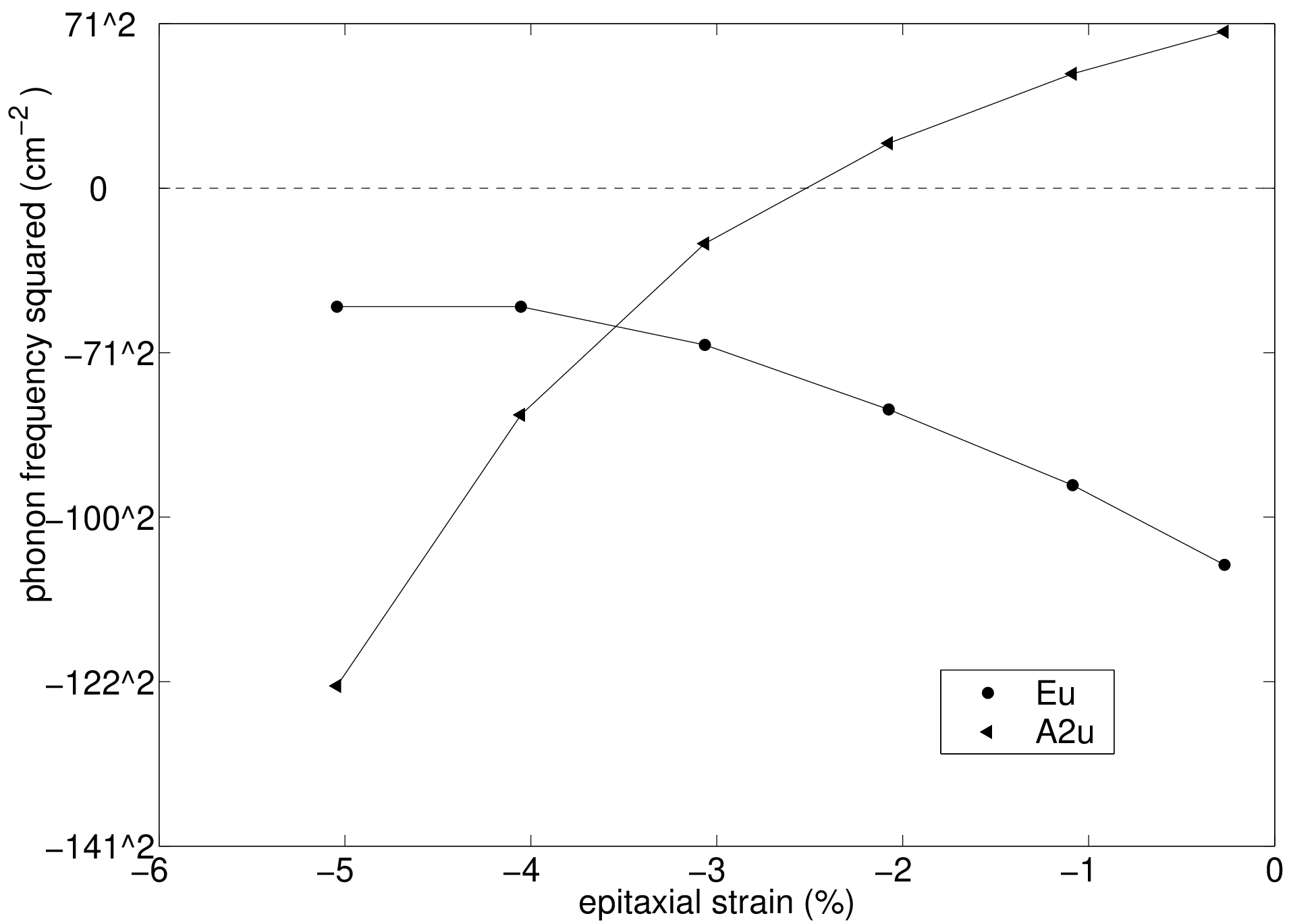}\\
\caption{\label{fig:phonons} 
Soft infrared-active phonon frequencies as a function of
in-plane compressive strain for the lowest frequency E$_u$ phonon and
lowest frequency A$_{2u}$ phonon of the high-symmetry, paraelectric $I4/mmm$
reference structure.}
\end{figure}

With this in mind we consider again the high-symmetry, $I4/mmm$ RP structure 
and explore the effect of epitaxial strain on the low-frequency infrared-active 
modes. We impose epitaxial strain by constraining the two basal primitive vectors 
of the bct lattice to an angle of 90 degrees and to a fixed equal length (i.e. 
corresponding to that of an implicit coherently matched square-lattice substrate).
In Fig.~\ref{fig:phonons} we show how the phonon frequencies for the lowest frequency 
E$_u$ and A$_{2u}$ phonons vary as a function of compressive epitaxial strain. 
We use the computed $a$ parameter of tetragonal ferroelectric PbTiO$_3$~\cite{my.pbtio3} 
as the reference strain (i.e. for 0$\%$ strain we fixed the in-plane lattice constant of 
Pb$_2$TiO$_4$ to that of ferroelectric PbTiO$_3$).~\cite{my.strain} For each value of 
fixed strain we again perform relaxation of the ions and $c$-axis. 

%%%%%%%%%%%%%%%
% Figure 2
%%%%%%%%%%%%%%

Referred to tetragonal PbTiO$_3$, the  ground-state $I4/mmm$ structure has an 
in-plane strain of approximately -0.3$\%$. For small compressive 
strains the phonon instabilities of epitaxial Pb$_2$TiO$_4$ are expected to be similar 
to those in the unconstrained structure.  This corresponds to the far right of the 
Fig.~\ref{fig:phonons}. It is evident that the lattice dynamics in this region of 
zero or small epitaxial strains are dominated by the largely unstable E$_u$ mode, 
as previously discussed. If we now increase the compressive epitaxial strain (i.e. 
from right-to-left) the E$_u$ mode stiffens while the A$_{2u}$ mode softens considerably 
and becomes unstable at $\approx$ -2.5$\%$ strain. Fig.~\ref{fig:phonons} shows 
that for large compressive strains (-4 to -5$\%$), the highly unstable A$_{2u}$ should dominate the 
lattice dynamics while for intermediate strain values, both an A$_{2u}$ mode and an 
E$_u$ mode are unstable and comparable in value. This behavior with strain can be 
simply understood as arising from volume effects. As we increase the in-plane compressive 
strain, the effective volume in which the E$_u$ mode (polarized in-plane, perpendicular 
to the $c$-axis) vibrates decreases. This increases the short-range repulsive forces, 
thereby stiffening the force constant.~\cite{ghosez.EUlett.96} In contrast, the effective 
volume of the A$_{2u}$ mode (polarized parallel to the $c$-axis) increases with increasing 
compressive strain leading to a softening of the force constant.

\begin{figure}[t]
\includegraphics[scale=0.25]{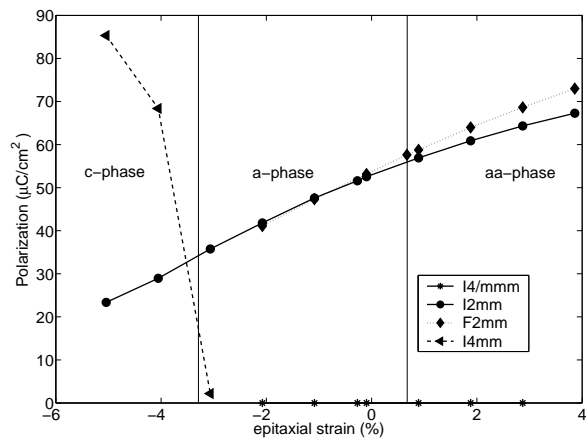}\\
\caption{\label{fig:polar}
Polarization along [001] ($I4mm$), 
[100] ($I2mm$), and [110] ($F2mm$) as a function of epitaxial  strain.}
\end{figure}

\begin{figure}[t]
\includegraphics[scale=0.25]{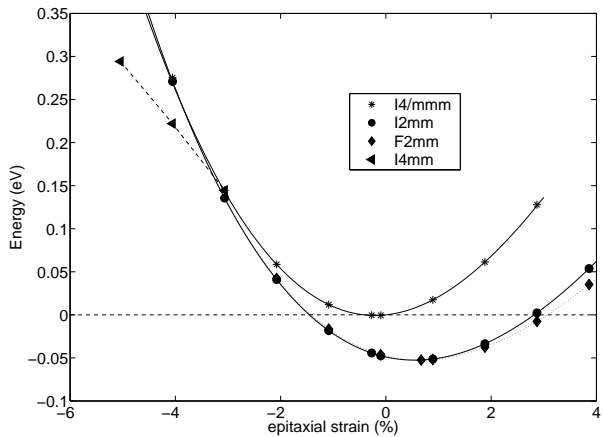}\\
\caption{\label{fig:energy}
Energy (per formula unit) as a function of epitaxial strain for space groups $I4/mmm$ (paraelectric),
$I2mm$ ($a$-phase), $F2mm$ ($aa$-phase), and $I4mm$ ($c$-phase).}
\end{figure}

Next we use these phonon instabilities (Fig.~\ref{fig:phonons}) as a guide to 
search for additional epitaxially stabilized ferroelectric structures.  At each 
value of strain we first freeze-in separately the real space eigendisplacement 
pattern corresponding to the A$_{2u}$ mode ($I4mm$), the E$_u$ [100] mode ($I2mm$), 
the E$_u$ [110] mode ($F2mm$), 
and both the A$_{2u}$ and the E$_u$ [100] modes ($Cm$). Then we relax all ions and the $c$-axis 
lattice parameter while keeping the in-plane lattice parameters fixed. Finally we calculate 
the total energy, Fig.~\ref{fig:energy}, and the polarization, Fig.~\ref{fig:polar}, of the 
resultant structures as a function of epitaxial strain.

%%%%%%%%%%%%%%%
% Figure 3
%%%%%%%%%%%%%%

As shown in Fig.~\ref{fig:polar}, RP Pb$_2$TiO$_4$ undergoes a series of
structural transitions with epitaxial strain. Over the range of slightly 
tensile to compressive strains, Pb$_2$TiO$_4$ forms in the $I2mm$ space 
group. Following the convention appearing in the literature,~\cite{pertsev} 
we refer to this phase as the $a$-phase. The polarization in this phase 
varies from 34$\mu$C/cm$^2$ at -3.3$\%$ strain to 56$\mu$C/cm$^2$ at +0.7$\%$ 
strain. The minimum energy structure in the $a$-phase occurs at a tensile 
strain of +0.55$\%$, corresponding to the ground state structure discussed 
above. As the compressive strain increases, the energy of the $I2mm$ 
structure approaches that of the paraelectric $I4/mmm$ structure, as shown in 
Fig.~\ref{fig:energy}, while remaining about 3 meV lower for the values of strain that we 
considered. This is consistent with the leveling off of the E$_u$ phonon as shown in 
Fig.~\ref{fig:phonons} and explains why the polarization of the $I2mm$ structure remains 
nonzero for large compressive strains. A transition from the $a$-phase, with the polarization 
in-plane, to a phase with the polarization along the $c$-axis, i.e. from $I2mm$ to $I4mm$, 
occurs for large compressive stains as anticipated from the phonon instabilities of the 
$I4/mmm$ structure. This occurs at $\approx$-3.3$\%$ strain. We refer to this $I4mm$ phase 
as the $c$-phase.  The polarization in the $c$-phase approaches 70$\mu$C/cm$^2$ at -4.0$\%$ 
strain and continues to increase. In the range where the two unstable mode frequencies cross, 
we considered the possibility that coupling between the two modes could lead to additional 
ferroelectric structures. However, relaxing the structures in the low symmetry space group 
$Cm$ always yielded one of the two higher symmetry structures ($a$ or $c$ phase, depending 
on the value of the misfit strain). Thus the transition from $a$ to $c$ with increasingly 
compressive in-plane strain appears to be first-order.~\cite{my.mono} 
Finally, for large enough tensile 
strains (greater than $\approx$0.7$\%$) the $F2mm$ structure becomes lower in energy than 
that of the $I2mm$ structure. The polarization of this $aa$-phase is comparable to that 
of the $a$-phase while the minimum energy structure occurs at a slightly more positive 
strain of +0.7$\%$ strain. The point at which the energy curves for the $a$ and $aa$ phases 
cross is of particular interest, as the in-plane polarization is nearly isotropic.
The free rotation of the polarization might result in some unexpected interesting physical 
properties.

%%%%%%%%%%%%%%%
%Figure 4
%%%%%%%%%%%%%%

%%%%%%%%%%%%%%%%%%%%%%%%%%%%%%%%%%%%%%%%%%%%%%%%%%%%%%%%%%%%%%%%%%%%%%%%%%%%%%%%%
%SUMMARY
%%%%%%%%%%%%%%%%%%%%%%%%%%%%%%%%%%%%%%%%%%%%%%%%%%%%%%%%%%%%%%%%%%%%%%%%%%%%%%%%%
%\section{Summary}
%\label{sec:summary}
Compounds unstable in the RP structure at atmospheric pressures have been 
synthesized under high-pressure conditions, e.g. polycrystalline
Ba$_2$RuO$_4$.~\cite{kafalas.jssc.72} This route seems 
promising to synthesize bulk RP Pb$_2$TiO$_4$. Further, we suggest synthesis of
non-bulk phases of this material through the use of epitaxial 
stabilization.~\cite{schlom.matsci.01,gorbenko.chemmat.02} In fact, this method
has proven quite successful for stabilizing a variety of oxide thin films in the RP
structure. One example is the higher order members ($n>3$) of the 
Sr$_{n+1}$Ti$_n$O$_{3n+1}$ series where bulk phases are known to be 
unstable.~\cite{haeni.apl.01} Another example has been the low pressure synthesis 
of Ba$_2$RuO$_4$ in the RP structure.~\cite{jia.apl.99}  The growth of thin films of 
Pb$_2$TiO$_4$ would provide a means to realize the interesting behavior of this material with 
epitaxial strain. 

Initially, we asked the question whether Pb$_2$TiO$_4$ in the Ruddlesden-Popper
structure would be a ferroelectric. Using first-principles DFT calculations, we 
have seen that indeed it does display a ferroelectric instability. We have argued 
that if Pb$_2$TiO$_4$ could be made in the RP structure (bulk or thin films) it 
would undergo a ferroelectric structural transition to the orthorhombic $a$-phase 
with a spontaneous polarization comparable to that of bulk PbTiO$_3$. 

%\acknowledgments
The authors would like to acknowledge many valuable discussions with Darrell Schlom.
We also thank David Singh for suggesting possible growth techniques at APS 2004. 
This work was supported by NSF-NIRT Grant No. DMR-0103354.


\begin{thebibliography}{10}
\bibitem{my.Ca3Ti2O7}
The $n$=2 RP Ca$_3$Ti$_2$O$_7$ crystallizes in a polar space group
although ferroelectricity has not been reported.
\bibitem{my.mono}
To definitively say whether the $Cm$ structure can be realized the 
monoclinic angle would have to be allowed to vary independently of the $c$-axis 
during cell relaxations, which was not done here for computational simplicity.
\end{thebibliography}
\end{document}